\documentclass[12pt]{article}  
\usepackage{color,amssymb,amsthm,amsmath,amsxtra,amsfonts,breqn,
fullpage,graphicx,hyperref,float,appendix,mathtools,eqparbox, 
lscape,pdfpages,comment}

\usepackage[latin1]{inputenc}
\usepackage{tikz}
\usetikzlibrary{shapes,arrows}

\usepackage{fancyhdr}
\usepackage{xcolor}
\definecolor{dark-red}{rgb}{0.4,0.15,0.15}
\definecolor{dark-blue}{rgb}{0,0,0.45}
\hypersetup{colorlinks, linkcolor={dark-blue},citecolor={blue}, urlcolor={blue}}

\numberwithin{equation}{section}
\usepackage{pdflscape}

\definecolor{gold}{rgb}{0.85,0.66,0.0}
\def\todo#1{\textcolor{red}{\textbf{**** TODO -- #1 ****}}}

\theoremstyle{definition}
\newtheorem{example}{Example}[section]
\newtheorem{definition}{Definition}

\usepackage{etoolbox}
\makeatletter
\providecommand{\institute}[1]{
  \apptocmd{\@author}{\end{tabular}
    \par
    \begin{tabular}[t]{c}
    #1}{}{}
}
\makeatother

\usepackage[ddmmyyyy,hhmmss]{datetime}

\begin{document}

\title{A Simple Method for Computing Some Pseudo-Elliptic Integrals in Terms of Elementary Functions}
\author{Sam Blake}
\institute{\textit{The University of Melbourne}}
\date{DRAFT version: \today:\currenttime}
\maketitle

\begin{abstract}
We introduce a method for computing some pseudo-elliptic integrals in terms of elementary 
functions. The method is simple and fast in comparison to the algebraic case of the 
Risch-Trager-Bronstein algorithm \cite{Risch1969}\cite{Risch1970}\cite{Trager1984}\cite{Bronstein1990}. 
This method can quickly solve many pseudo-elliptic integrals, which other well-known 
computer algebra systems either fail, return an answer in terms of special functions, 
or require more than 20 seconds of computing time. Randomised tests showed our method 
solved 73.4\% of the integrals that could be solved with the best implementation of the Risch-Trager-Bronstein 
algorithm. Unlike the symbolic integration algorithms of Risch \cite{Risch1969}\cite{Risch1970}, 
Davenport \cite{Davenport1979}, Trager \cite{Trager1984}, Bronstein \cite{Bronstein1990} and 
Miller \cite{Miller2012}; our method is not a decision process. The implementation of this method 
is less than 200 lines of Mathematica code and can be easily ported to other CAS that can solve systems 
of polynomial equations.
\end{abstract}





\section{Introduction}
The problem of finding elementary solutions to integrals of algebraic functions has 
challenged mathematicians for centuries. In 1905, Hardy conjectured that the problem may be 
unsolvable \cite{Hardy1916}. Sophisticated algorithms have been developed, including 
the famed Risch algorithm \cite{Risch1969}\cite{Risch1970} and its modern variants by 
Davenport \cite{Davenport1979}, Trager \cite{Trager1984}, Bronstein \cite{Bronstein1990} 
and Miller \cite{Miller2012}. However, it is known that the implementation of these 
algorithms are highly complex and sometimes fail, are incomplete or hang \cite{fricas_risch_status}. \\



We will be describing a seemingly new method for computing pseudo-elliptic integrals. 
\begin{definition}
For our purposes, a pseudo-elliptic integral is of the form 
$$\int \frac{p(x)}{q(x)}r(x)^{n/m}dx,$$
where $n,m \in \mathbb{Z}$, $p(x),q(x),r(x) \in \mathbb{Z}[x]$, $\deg_x(r(x))>2$, 
$\gcd(n,m)=1$, and possesses a solution in terms of elementary functions. 
\end{definition} 

Before we describe our method, we begin with some background and motivating examples. \\

The \textit{derivative divides} method is a substitution method that finds all composite 
functions, $u=g(x)$, of the integrand $f(x)$, and tests if $f(x)$ divided by the derivative of 
$u$ is independent of $x$ after the substitution of $u=g(x)$. In other words, up to a constant
factor the derivative divides method simplifies integrals of the form $\int f(g(x))g'(x)dx$ to 
$\int f(u) \, du$. The following integral illustrates the derivative divides method 
$$\int x^2\sqrt{1+x^3}dx=\int \frac{\sqrt{u}}{3} \, du,$$ 
where $u=1+x^3$ and $du=3x^2dx$. This method was first implemented in Moses symbolic integrator, 
SIN, in 1967 \cite{Moses1967} and is used in some CAS prior to calling more advanced 
algorithms \cite[pp. 473-474]{Geddes1992}.\\

A slightly more difficult example where the derivative divides method fails is 
$$\int \sqrt{x-1+\sqrt{x-1}} \, dx.$$
In this case we make the substitution $u=\sqrt{x-1}$, then $2udu=dx$. Furthermore, we need to 
express $x-1$ in terms of $u$, which is $u^2=x-1$. Then the integral becomes
$$\int 2u\sqrt{u^2+u}\,du.$$
While this integral was more difficult than the previous example, we could still 
pick a composite function, $\sqrt{x-1}$, of our integrand, $\sqrt{x-1+\sqrt{x-1}},$ to use 
as our $u$ substitution. In the following well-known example, this is not immediately possible
$$\int \frac{x^2-1}{\left(x^2+1\right)\sqrt{1+x^4}} \, dx.$$
A common approach to solve this integral is rearranging the integrand into
$$\int \frac{x^2\left(1-1\left/x^2\right.\right)}{x(x+1/x)\sqrt{x^2\left(x^2+1\left/x^2\right.\right)}} \, dx = 
\int \frac{1-1\left/x^2\right.}{(x+1/x)\sqrt{(x+1/x)^2-2}} \, dx. $$ 
Then the substitution $u=x+1/x$, $du=\left(1-1\left/x^2\right.\right)dx$ yields the integral
$$\int \frac{du}{u\sqrt{u^2-2}},$$
which, in general, may be transformed into a rational function using the Euler substitution \cite{Euler}.\\

While heuristics like algebraic rearranging and integration by substitution are simple, if a 
computer algebra system (CAS) is available, why would you use such methods instead of the 
algebraic case of the Risch-Trager-Bronstein algorithm? By way of explanation, consider 
the following integral 
\begin{equation*}
\int\frac{2 a^2 x^4-b^2}{\left(a^2 x^8-b^2\right)\sqrt[4]{a^2 x^4-b^2}}dx,
\end{equation*}
where $a,b\in\mathbb{C}$. Using the substitutions $u=\left(x^4-b^2\right)/x^4$, and 
$t^4=u+a^2-1$, this integral is reduced to integrating a rational function, 
\begin{equation*}
-\int\frac{t^2 \left(t^4+a^2\right)}{\left(t^4-a^2-a b\right) \left(t^4-a^2+a b\right)}dt,
\end{equation*}
and the solution to our original integral is 
\begin{multline*}
\frac{1}{4b \sqrt[4]{a} \sqrt[4]{a-b}} \left( 
\left(b-2a\right)\tan ^{-1}\left(\frac{\sqrt[4]{a}\sqrt[4]{a-b}\,x}{\sqrt[4]{a^2x^4-b^2}}\right) + 
\left(b-2a\right)\tanh ^{-1}\left(\frac{\sqrt[4]{a}\sqrt[4]{a-b}\,x}{\sqrt[4]{a^2 x^4-b^2}}\right)\right. +\\ 
\left. \left(2a+b\right)\tan^{-1}\left(\frac{\sqrt[4]{a}\sqrt[4]{a+b}\,x}{\sqrt[4]{a^2 x^4-b^2}}\right) + 
\left(2a+b\right) \tanh ^{-1}\left(\frac{\sqrt[4]{a}\sqrt[4]{a+b}\,x}{\sqrt[4]{a^2 x^4-b^2}}\right)
\right).
\end{multline*}
The symbolic integration routines of Maple (2018.1) and Mathematica (12.1.0) cannot compute this integral. Both 
AXIOM and FriCAS seemingly hang, with no response after 2 hours. As the integration of rational functions
(via Hermite reduction and the Lazard-Rioboo-Trager algorithm \cite{Bronstein1997}) is significantly faster than the 
Risch-Trager-Bronstein algorithm for algebraic integration, if we can quickly determine when a rational 
substitution (or more precisely a Laurent polynomial substitution) will reduce the integrand to one  
of the forms $\mathbb{Q}(x)$, $\mathbb{Q}(x, (a\,x+b)^{n/m})$, or $\mathbb{Q}(x,(a\,x^2+b\,x+c)^{n/2})$, 
then we would have a reasonable alternative to the heavy algebraic computations required by the 
Risch-Trager-Bronstein algorithm. This approach becomes more attractive when you consider that in major 
computer algebra systems the algebraic case (and the mixed algebraic-transcendental case) of the 
Risch-Trager-Bronstein algorithm is either partially implemented, not implemented, or contains computational 
bottlenecks that result in long computations. \\

\section{A method for solving some pseudo-elliptic integrals}
Following on from our example integral $\int \frac{x^2-1}{\left(x^2+1\right)\sqrt{1+x^4}} \, dx$, 
where making the substitution $u=x+1/x$ resulted in the integral $\int \frac{du}{u\sqrt{u^2-2}}$. 
We would like to generalise this method, however the difficulty was in the choice of the algebraic 
manipulation to the form 
$\int\frac{1-1\left/x^2\right.}{(x+1/x)\sqrt{(x+1/x)^2-2}} \, dx$ in order to discover a rational 
substitution, which simplifies the integral. Consequently our approach does not directly rely 
on such an algebraic manipulation of the integrand. \\

Our method attempts to parameterise constants $a_0, a_1, a_2$, polynomials $a(u), b(u)$ and a 
Laurent polynomial substitution of the form 
$$u = \frac{s(x)}{x^k}$$ 
such that
\begin{equation}
\int \frac{p(x)}{q(x)}r(x)^{n/m}dx = \int \frac{a(u)}{b(u)}\left(a_2u^2+a_1u+a_0\right)^{n/m}du,\label{reduction}
\end{equation}
where $\deg_x(r(x))>2$ and $\gcd (n,m)=1$. Consequently, our method does not directly compute the 
integral, and requires a recursive call to an algebraic integrator\footnote{If such an integrator is not 
available then a reasonable implementation could call a lookup table of algebraic forms \cite{Prudnikov1986} 
followed by a rational function integrator \cite{Bronstein1997}.}. We note that a reduction to this form
does not guarantee an elementary solution (for example, when $m>2,a_2\ne0$ an elementary form 
is often not possible). \\

Our method is broken into two parts. The first part is computing the \textit{radicand part of 
the integral}, which is a parameterisation of $a_0,a_1,a_2$ and the $u$ 
substitution. The second part is computing the \textit{rational part of the integral}, which is 
a parameterisation of $a(u)$ and $b(u)$.\\

\textit{The radicand part of the integral.} Clearly, if we cannot parameterise the radicand $r(x)$ 
to the form $a_2u^2+a_1u+a_0$ for a given substitution, then we cannot find a 
parameterisation of (\ref{reduction}). Thus, we begin by computing the radicand part 
of the integral, which requires solving 
\begin{equation}
r(x) = \text{num}\left( a_2u^2 + a_1u + a_0 \right), \label{reduction_radicand}
\end{equation}
for the constants $a_0,a_1,a_2$ and the substitution $u = s(x)/x^k$. We do this by iterating over 
$0 < d \leq \text{deg}_x(r(x))$ such that $s(x) = \sum\limits_{i=0}^{d} c_i\,x^{i}$, where for each $d$ we 
iterate over $0 \leq h \leq \text{deg}_x(r(x))$, where $u = s(x)/x^h$ and $ h \mod m = 0$.  
Given a candidate $u$, we solve (\ref{reduction_radicand}) by equating coefficients of $x$ and solving the system of 
equations for the unknowns $a_0$, $a_1$, $a_2$, $c_0$, $c_1$, $\cdots$, $c_d$. If a solution 
(or multiple solutions) exists we move to computing the rational part of the integral, otherwise 
we move onto the next radicand or candidate substitution. If no solution exists to the radicand 
part of the integral for any candidate $u$-substitutions, then our method fails to 
compute the integral.\\

\textit{The rational part of the integral.} Given the substitution and solution set of 
the radicand part of the integral, we now look to solve the rational part of the integral, 
which is given by 
\begin{equation}
\frac{p(x)}{q(x)} = \frac{a(u(x))}{b(u(x))}\frac{u'(x)}{\text{den}\left(a_2u(x)^2+a_1u(x)+a_0\right)^{n/m}}, \label{reduction_rational}
\end{equation}
where $a_0$, $a_1$, $a_2$, $u(x)$ are known and $a(u)$, $b(u)$ are unknown. The degree bound estimate  
of $a(u)$ and $b(u)$ is given by $\mathcal{D} = \deg_x(u(x)) + \deg_x(u'(x)) + \max\left( \deg_x(p(x)), \deg_x(q(x)) \right)$. We solve 
(\ref{reduction_rational}) by increasing the degree, $d$, of $a(u)$ and $b(u)$ from 1 to the degree 
bound, $\mathcal{D}$, where for each iteration we solve
\begin{equation*}
p(x)\,b(u(x))\,\text{den}\left(a_2u(x)^2 + a_1u(x) + a_0\right)^{n/m} - q(x)\,a(u(x))\,u'(x) = 0,
\end{equation*}
where $a(u) = \sum\limits_{i=0}^d v_i u^i$, $b(u) = \sum\limits_{i=0}^d v_{d+i+1} u^i$, and 
$a(u(x))$, $b(u(x))$ are rational functions in $x$ after replacing $u$ with the candidate 
substitution. As before, we equate powers of $x$ and solve for the unknowns $v_0$, $v_1$, 
$\cdots$, $v_{2d+1}$. If a solution is found, then we have a complete solution to 
(\ref{reduction}) and we stop. Otherwise if we have iterated up to the degree bound, and 
iterated through all solution sets from the radicand part of the integral and we have not 
computed a solution, then the candidate substitution is rejected and must return to the 
radicand part of the integral to try the next substitution.

\begin{example}
We will apply the method detailed above to compute the following integral
$$\int \frac{\left(x^3-2\right) \sqrt{x^3-x^2+1}}{\left(x^3+1\right)^2} \, dx.$$ 

\noindent\textit{The radicand part of the integral}. We find the substitution 
$u=\left(c_1 x^3+c_0\right)/x^2$ yields a solution to the radicand part of the 
integral. As the degree of the radicand is odd, we must have $a_2=0$, so we have 
a linear radicand in $u$ as follows 
$$x^3-x^2+1 = \text{num}\left( a_1u + a_0 \right) = a_1 c_1 x^3 + a_0 x^2 + a_1 c_0.$$
Equating coefficients of powers of $x$, we have the following system of equations
\begin{align*}
 a_1 c_0 &= 1\\
 a_0 &= -1\\
 a_1 c_1 &= 1,
\end{align*}
which has the solution $a_0 = -1, a_1 = 1, c_0 = 1, c_1 = 1$. Thus, the radicand part of the 
integral is $u-1$, where $u=\left(1+x^3\right)/x^2$. \\

\noindent\textit{The rational part of the integral}. Now we see if a solution exists to the 
rational part of the integral. The degree bound on the solution to the rational part 
is 7. When the degree is 1, we have no solution. When the degree is 2, we have
$$\frac{a(u)}{b(u)}=\frac{v_2u^2+v_1u+v_0}{v_5u^2+v_4u+v_3}.$$
For the rational part, we are solving the following equation
$$\frac{\left(x^3-2\right)}{\left(x^3+1\right)^2} = 
	\left(\frac{v_2u^2+v_1u+v_0}{v_5u^2+v_4u+v_3}\right)\text{den}(u-1)^{-1/2}\,u'(x),$$
where $\text{den}(u-1)^{-1/2}=1/x$. After replacing $u$ with $\left(1+x^3\right)/x^2$ and 
$u'(x)$ with $\left(x^3-2\right)/x^3$ we have
$$\frac{x^3-2}{\left(x^3+1\right)^2}=
\frac{\left(x^3-2\right) \left(u^2 v_2+u v_1+v_0\right)}{x^4 \left(u^2 v_5+u v_4+v_3\right)}=
\frac{\left(x^3-2\right)\left(x^4 v_0+x^2 v_1+x^5 v_1+v_2+2 x^3 v_2+x^6 v_2\right)}{x^4 \left(x^4 v_3+x^2 v_4+x^5 v_4+v_5+2 x^3 v_5+x^6 v_5\right)},$$
which after clearing denominators is a polynomial equation in $x$, given by
\begin{multline*}
-v_2 x^{15}-v_1 x^{14}+\left(v_5-v_0\right) x^{13}+\left(v_4-2 v_2\right) x^{12}
+\left(v_3-v_1\right) x^{11}+\left(2 v_2-v_4\right) x^9+ \\
\left(3v_1-2 v_3\right) x^8+\left(3 v_0-3 v_5\right) x^7+
\left(8 v_2-2 v_4\right) x^6+5 v_1 x^5+\left(2 v_0-2 v_5\right) x^4+7 v_2 x^3+2 v_1 x^2+2 v_2=0,
\end{multline*}
which we solve for the undetermined coefficients $v_0,v_1,v_2,v_3,v_4,v_5$. Then equating coefficients of powers of $x$ yields the system of equations

\begin{align*}
 2 v_2&=0 \\
 2 v_1&=0 \\
 7 v_2&=0 \\
 2 v_0-2 v_5&=0 \\
 5 v_1&=0 \\
 8 v_2-2 v_4&=0 \\
 3 v_0-3 v_5&=0 \\
 3 v_1-2 v_3&=0 \\
 2 v_2-v_4&=0 \\
 -v_1+v_3&=0 \\
 -2 v_2+v_4&=0 \\
 -v_0+v_5&=0 \\
 -v_1&=0 \\
 -v_2&=0,
\end{align*}

which has the solution $v_0=v_5,v_1=0,v_2=0,v_3=0,v_4=0$. Thus, the rational part of the integral is
$$\frac{v_0}{v_0u^2}=\frac{1}{u^2}$$
and the integral is given by 
\begin{multline*}
\int \frac{\left(x^3-2\right) \sqrt{x^3-x^2+1}}{\left(x^3+1\right)^2} \, dx 
=\int \frac{\sqrt{u-1}}{u^2} \, du \\
=-\frac{\sqrt{u-1}}{u}+\tan ^{-1}\left(\sqrt{u-1}\right)
=-\frac{x\sqrt{x^3-x^2+1}}{x^3+1}+\tan ^{-1}\left(\frac{\sqrt{x^3-x^2+1}}{x}\right).\\
\end{multline*}
\end{example}

Our implementation in Mathematica took 0.085 seconds to compute this integral.

\section{A comparison with major CAS and algebraic integration packages}

We will compare our method with the Mathematica (12.1.0), Maple (2018.1), AXIOM (August 2014), 
REDUCE (5286, 1-Mar-20) with the algint package by James Davenport \cite{Davenport1979}, and 
FriCAS (1.3.6) computer algebra systems. We will also include 
in the comparison a table lookup package, Rubi (Rule-based integrator) \cite{Rich2018}, which 
has been ported to a number of computer algebra systems and compares favourably with most 
built--in integrators on a large suite of problems \cite{rubi_results}. We have also included 
an experimental algebraic integration package developed in Mathematica by Manuel Kauers 
\cite{Kauers2008}. Within this package we have replaced the calls to Singular in favour
of Mathematica's built--in Groebner basis routine.\\

We have included results from Maple twice. Once with a call of \texttt{int(integrand, x)} 
and once with \texttt{int(convert(integrand, RootOf),x)}. This is because the default 
behaviour of Maple is to not use the Risch-Trager-Bronstein integration 
algorithm for algebraic functions unless the radicals in the integrand are converted 
to the Maple \texttt{RootOf} notation \cite[pp. 16--23]{rybowicz1992}.\\

Our test suite is 191 integrals that can be found on github \cite{test_suite_github}. 
All the integrals in the suite have a solution in terms of elementary functions. \\

We will show the results from all the systems and packages for one integral from the test 
suite. It is intriguing to see the variety of forms for this integral. \\

\noindent Our method returns:
\small
\begin{equation*}
\int \frac{\left(x^4-1\right) \sqrt{x^4+1}}{x^8+1} \, dx= 
-\frac{1}{2 \sqrt[4]{2}}\tan ^{-1}\left(\frac{\sqrt[4]{2} x}{\sqrt{x^4+1}}\right) - 
\frac{1}{2 \sqrt[4]{2}}\tanh^{-1}\left(\frac{\sqrt[4]{2} x}{\sqrt{x^4+1}}\right)
\end{equation*}
\normalsize
FriCAS returns:
\small
\begin{align*}
& \int \frac{\left(x^4-1\right) \sqrt{x^4+1}}{x^8+1} \, dx = \\
& \frac{1}{8 \sqrt[4]{2}}\log\left(\frac{1}{x^8+1}\left(4 x^6+4 x^2+\sqrt{2} \left(x^8+4x^4+1\right)
-\sqrt{x^4+1} \left(2^{3/4} \left(2 x^5+2 x\right)+4 \sqrt[4]{2} x^3\right)\right)\right) - \\
&\frac{1}{8 \sqrt[4]{2}}\log \left(\frac{-1}{x^8+1}\left(4x^6+4 x^2+\sqrt{2} \left(x^8+4 x^4+1\right)+
\sqrt{x^4+1} \left(2^{3/4} \left(2 x^5+2 x\right)+4 \sqrt[4]{2} x^3\right)\right)\right) + \\
& \frac{1}{2\sqrt[4]{2}}\tan ^{-1}\left(\frac{-4 x^6-4 x^2+\sqrt{2} \left(x^8+4 x^4+1\right)}{\sqrt{2} \left(-x^8-1\right)+\sqrt{x^4+1} \left(2^{3/4} \left(2x^5+2 x\right)-4 \sqrt[4]{2} x^3\right)}\right)
\end{align*}
\normalsize

\bigskip

\noindent AXIOM returns the integral unevaluated, which is a claim that the integral is not elementary \cite[pp. 120]{Daly2005}. \\ 

\noindent Kauers' method returns: 
\small
$$\int \frac{\left(x^4-1\right) \sqrt{x^4+1}}{x^8+1} \, dx = 
\sum_{512 \alpha ^4-1=0}\alpha  \log \left(4 \alpha  \sqrt{x^4+1}-x\right)$$
\normalsize
Maple (default) returns:
\small
$$\int \frac{\left(x^4-1\right) \sqrt{x^4+1}}{x^8+1} \, dx = 
\frac{1}{2 \sqrt[4]{2}} \tan ^{-1}\left(\frac{\sqrt{x^4+1}}{\sqrt[4]{2} x}\right) - 
\frac{1}{4 \sqrt[4]{2}} \log\left(\frac{\frac{\sqrt{x^4+1}}{\sqrt{2} x}+\frac{1}{\sqrt[4]{2}}}{\frac{\sqrt{x^4+1}}{\sqrt{2} x}-\frac{1}{\sqrt[4]{2}}}\right)$$
\normalsize
Maple (with the \texttt{RootOf} conversion) returns:
\small
\begin{multline*}
\int \frac{\left(x^4-1\right) \sqrt{x^4+1}}{x^8+1} \, dx = 
\frac{1}{4 \sqrt[4]{2}}\log \left(\frac{2\times2^{3/4} x^4-8 \sqrt{x^4+1} x+4 \sqrt[4]{2} x^2+2\times2^{3/4}}{-2 x^4+2 \sqrt{2} x^2-2}\right) + \\
\frac{i}{4 \sqrt[4]{2}}\log \left(\frac{2\times2^{3/4} i x^4-8 \sqrt{x^4+1} x-4 i \sqrt[4]{2} x^2+2\times2^{3/4} i}{2 x^4+2 \sqrt{2} x^2+2}\right)
\end{multline*}
\normalsize
Mathematica returns:
\small
\begin{multline*}
\int \frac{\left(x^4-1\right) \sqrt{x^4+1}}{x^8+1} \, dx = \frac{1}{2} \sqrt[4]{-1} \left(-2 F\left(\left.i \sinh ^{-1}\left(\sqrt[4]{-1} x\right)\right|-1\right) \right. + \\
\Pi\left(-\sqrt[4]{-1};\left.i \sinh ^{-1}\left(\sqrt[4]{-1} x\right)\right|-1\right) + 
\Pi \left(\sqrt[4]{-1};\left.i \sinh ^{-1}\left(\sqrt[4]{-1} x\right)\right|-1\right) + \\
\left. \Pi\left(-(-1)^{3/4};\left.i \sinh ^{-1}\left(\sqrt[4]{-1} x\right)\right|-1\right)+\Pi \left((-1)^{3/4};\left.i \sinh ^{-1}\left(\sqrt[4]{-1} x\right)\right|-1\right)\right)
\end{multline*}
\normalsize
where \href{https://reference.wolfram.com/language/ref/EllipticF.html}{\textit{F}} and 
\href{https://reference.wolfram.com/language/ref/EllipticPi.html}{$\Pi$} are the incomplete 
elliptic integrals of the first and third kind as defined in Mathematica. \\

\noindent REDUCE (using the algint package) returns:
\small
$$\int \frac{\left(x^4-1\right) \sqrt{x^4+1}}{x^8+1} \, dx = 
\int \frac{x^4 \sqrt{x^4+1}}{x^8+1} \, dx-\int \frac{\sqrt{x^4+1}}{x^8+1} \, dx$$
\normalsize
Rubi returns\footnote{After posting a preprint of this comparison on the 
\href{https://groups.google.com/forum/\#!forum/sci.math.symbolic}{sci.math.symbolic}
newsgroup, Albert Rich (the creator of Rubi) devised a general rule for integrals of this 
form which will be included in the next release of Rubi: %
$$\int\frac{\left(f+g x^4\right)\sqrt{d+e x^4}}{a+b x^4+c x^8} dx = 
\frac{e^2 f}{2 c d \sqrt[4]{2 d e-\frac{b e^2}{c}}} \tan ^{-1}\left(\frac{x \sqrt[4]{2 d e-\frac{b e^2}{c}}}{\sqrt{d+e x^4}}\right) 
	+\frac{e^2 f}{2 c d \sqrt[4]{2 d e-\frac{b e^2}{c}}} \tanh ^{-1}\left(\frac{x \sqrt[4]{2 d e-\frac{be^2}{c}}}{\sqrt{d+e x^4}}\right),$$ when $e f+d g=0$ and $c d^2-a e^2=0$.}:
\small
\begin{multline*}
\int \frac{\left(x^4-1\right) \sqrt{x^4+1}}{x^8+1} \, dx=
-\frac{\tan ^{-1}\left(\frac{\sqrt[4]{2} x}{\sqrt{x^4+1}}\right)}{2 \sqrt[4]{2}}-
\frac{\tanh^{-1}\left(\frac{\sqrt[4]{2} x}{\sqrt{x^4+1}}\right)}{2 \sqrt[4]{2}} + \\
\frac{\left(x^2+1\right) \sqrt{\frac{x^4+1}{\left(x^2+1\right)^2}} F\left(2\tan ^{-1}(x)|\frac{1}{2}\right)}{2 \sqrt{x^4+1}} + 
\frac{\left((-1-i)-i \sqrt{2}\right) \left(x^2+1\right) \sqrt{\frac{x^4+1}{\left(x^2+1\right)^2}}
F\left(2 \tan ^{-1}(x)|\frac{1}{2}\right)}{8 \sqrt{x^4+1}} + \\
\frac{\left(\sqrt{2}+(-1+i)\right) i \left(x^2+1\right) \sqrt{\frac{x^4+1}{\left(x^2+1\right)^2}}
F\left(2 \tan ^{-1}(x)|\frac{1}{2}\right)}{8 \sqrt{x^4+1}} + \\
\frac{\left(\sqrt{2}+(1+i)\right) i \left(x^2+1\right) \sqrt{\frac{x^4+1}{\left(x^2+1\right)^2}}
F\left(2 \tan ^{-1}(x)|\frac{1}{2}\right)}{8 \sqrt{x^4+1}} - \\
\frac{\left(\frac{1}{8}-\frac{i}{8}\right) \left(1+(-1)^{3/4}\right) \left(x^2+1\right)
\sqrt{\frac{x^4+1}{\left(x^2+1\right)^2}} F\left(2 \tan ^{-1}(x)|\frac{1}{2}\right)}{\sqrt{x^4+1}}
\end{multline*}
\normalsize
where \href{https://reference.wolfram.com/language/ref/EllipticF.html}{\textit{F}} is the incomplete 
elliptic integral of the first kind as defined in Mathematica. Noting that the last 5 terms 
in the result from Rubi sum to zero, we obtain the same result as our method. \\

The table below summarises the comparison between all systems on the test suite of integrals. The integrals 
in this suite were created in such a way that our method described in this paper should solve. Similarly, 
computer algebra systems containing an implementation of the algebraic case of the Risch-Trager-Bronstein 
algorithm should solve all these integrals. 

\begin{table}[H]
\centering
\caption{\small A comparison of our method (listed as ``new'') with major CAS and algebraic integration packages. The 
median time excludes integrals that timed-out. The tests were ran on a 2018 Macbook Pro 
with a 2.2GHz i7 and 16GB of RAM. For AXIOM, FriCAS and REDUCE we did not find a built-in 
routine to compute the leaf count. \normalsize}
{\small
\begin{tabular}{c|c|c|c|c|c|c}
	\begin{tabular}{@{}c@{}} \text{CAS} \\ \text{/package} \end{tabular} & 
	\begin{tabular}{@{}c@{}} \text{Elementary} \\ \text{forms [\%]} \end{tabular} &
	\begin{tabular}{@{}c@{}} \text{Contains} \\ \text{$\int dx$ [\%]} \end{tabular} &
	\begin{tabular}{@{}c@{}} \text{Contains} \\ \text{special} \\ \text{functions [\%]} \end{tabular} &
	\begin{tabular}{@{}c@{}} \text{Timed-out} \\ \text{($>$20s) [\%]} \end{tabular} &
	\begin{tabular}{@{}c@{}} \text{Median} \\ \text{time [s]} \end{tabular} &
	\begin{tabular}{@{}c@{}} \text{Median} \\ \text{expression size} \\ \text{- string length} \end{tabular} \\
\hline
 \text{new} & 100.0 & 0.0 & 0.0 & 0.0 & 0.26 & 81\:--\:118 \\
\hline
 \text{Maple} (\texttt{RootOf}) & 91.6 & 2.0 & 0.0 & 6.3 & 4.32 & 136\:--\:238 \\
\hline
 \text{FriCAS} & 75.4 & 2.1 & 0.0 & 18.3 & 0.38 & .\text{  }\:--\:186 \\
\hline
 \text{Kauers} & 62.6 & 7.9 & 0.0 & 29.4 & 0.40 & 88\:--\:116 \\
\hline
\text{AXIOM} & 48.7 & 31.9 & 0.0 & 19.4 & 0.42 & .\text{  }\:--\:118 \\
\hline
 \text{Rubi} & 13.7 & 60.0 & 16.3 & 10.0 & 0.34 & 244\:--\:405 \\
\hline
 \text{Maple} & 11.6 & 55.3 & 33.2 & 0.0 & 0.34 & 436\:--\:943 \\
\hline
 \text{Mathematica} & 9.5 & 44.7 & 43.2 & 2.63 & 1.21 & 574\:--\:1019 \\
\hline
 \text{REDUCE} (\text{algint}) & 6.3 & 65.3 & 0.0 & 28.4 & 1.16 & . \\
\end{tabular}
}
\end{table}

Our method can quickly solve all these integrals, however this is not surprising as the suite 
of integrals was designed so that a Laurent polynomial substitution would reduce the integral 
to a simple elementary integral. What is interesting in the results from this suite is the 
performance of the computer algebra systems. Our naive expectation was the computer algebra 
systems with an implementation of the Risch-Trager-Bronstein algorithm would solve these 
integrals. \\

We would like to estimate the percentage of algebraic integrals which may be solved with 
our method. We do this with a random search for integrals with elementary solutions, using 
Maple (with the \texttt{RootOf} conversion), as it was the best performing CAS in 
our first suite of integrals. The suite of 384 integrals are available on 
github \cite{test_suite_github_2}. Our method could compute 73.4\% of these integrals in 
a median time of 0.23 seconds. \\

\section{Conclusions}

We have shown that we can efficiently solve some pseudo-elliptic integrals in terms of 
elementary functions. Our method compares favourably with major CAS and algebraic 
integration packages.  \\

The computational burden of our method is low, as the core routine requires 
solving multiple systems of polynomial equations, and consequently our method should be tried before the 
more computationally expensive algorithms of Trager \cite{Trager1979}\cite{Trager1984}, Bronstein 
\cite{Bronstein1990}, Kauers \cite{Kauers2008}, or Miller \cite{Miller2012}.\\

Our method, relative to the algebraic case of the Risch-Bronstein-Trager algorithm, is very simple to 
implement. Our exemplar implementation in Mathematica is only a couple of hundred lines of code and 
relies heavily on \texttt{SolveAlways} for solving systems of polynomial equations with undetermined 
coefficients. The implementation is available on github \cite{algebraic_github}. 

\section{Acknowledgements}
I would like to thank Albert Rich and David Stoutemyer for their detailed comments and suggestions; and 
thank Daniel Lichtblau for an explanation of block ordering of Groebner basis in Mathematica. I am 
grateful for the use of the Spartan high performance computing system at The University of 
Melbourne \cite{lafayette2016}. 

\bibliographystyle{abbrv}

\end{document}